\documentstyle[12pt]{article}

\begin{document}
\begin{titlepage}
\begin{center}

February 27, 2002     \hfill    LBNL-48917-REV \\

\vskip .3in

{\large \bf The Basis Problem in Many-Worlds Theories}
\footnote{This work is supported in part by the Director, Office of Science, 
Office of High Energy and Nuclear Physics, Division of High Energy Physics, 
of the U.S. Department of Energy under Contract DE-AC03-76SF00098}

\vskip .30in
Henry P. Stapp\\
{\em Lawrence Berkeley National Laboratory\\
      University of California\\
    Berkeley, California 94720}
\end{center}

\vskip .3in

\begin{abstract}

It is emphasized that a many-worlds interpretation of quantum theory exists 
only to the extent that the associated basis problem is solved. The core
basis problem is that the robust enduring states specified by environmental
decoherence effects are essentially Gaussian wave packets that form continua 
of non-orthogonal states. Hence they are not a discrete set of orthogonal 
basis states to which finite probabilities can be assigned by the usual rules.
The natural way to get an orthogonal basis without going outside 
the Schroedinger dynamics is to use the eigenstates of the reduced density 
matrix, and this idea is the basis of some recent attempts by many-worlds 
proponents to solve the basis problem. But these eigenstates do not enjoy the 
locality and quasi-classicality properties of the states defined by 
environmental decoherence effects, and hence are not satisfactory preferred 
basis states. This core problem needs to be addressed and resolved
before a many-worlds-type interpretation can be said to exist.  
\end{abstract}

\end{titlepage}

\newpage
\renewcommand{\thepage}{\arabic{page}}
\setcounter{page}{1}

{\bf 1. Introduction.}

The basis problem of the many-worlds interpretation can be efficiently
introduced by quoting a few exerpts from the entry  "Everett's Relative-State 
Formulation of Quantum Mechanics" in the Stanford Encyclopedia of Philosophy. 
That entry is available online\cite{bar}, and elaborates on the points 
summarized in the following passages: 

\begin{quote}
  Everett's relative-state formulation of quantum mechanics is an
  attempt to solve the measurement problem by dropping the collapse
  dynamics from the standard von Neumann-Dirac theory of quantum
  mechanics.  The main problem with Everett's theory is that it is not
  at all clear how it is supposed to work.  In particular, while it is
  clear that he wanted to explain why we get determinate measurement
  results in the context of his theory, it is unclear how he intended to
  do this.  There have been many attempts to reconstruct Everett's
  no-collapse theory in order to account for the apparent
  determinateness of measurement outcomes.  These attempts have led to
  such formulations of quantum mechanics as the many-worlds, many-minds,
  many-histories, and relative-fact theories.  Each of these captures
  part of what Everett claimed for his theory, but each also encounters
  problems.\\
  \ldots\ \\
  Everett's no-collapse formulation of quantum mechanics was a reaction
  to problems that arise in the standard von Neumann-Dirac collapse
  theory.  Everett's proposal was to drop the collapse postulate from
  the standard theory, then deduce the empirical predictions of the
  standard theory as the subjective experiences of observers who are
  themselves treated as physical systems described by his theory.  It
  is, however, unclear precisely how Everett intended for this to work. 
  Consequently, there have been many, mutually incompatible, attempts 
  at trying to explain what he in fact had in mind.\\
  \ldots\ \\
  This is what Everett says: "We shall be able to introduce into [the
  relative-state theory] systems which represent observers.  Such
  systems can be conceived as automatically functioning machines
  (servomechanisms) possessing recording devices (memory) and which are
  capable of responding to their environment.  The behavior of these
  observers shall always be treated within the framework of wave
  mechanics.  Furthermore, we shall deduce the probabilistic assertions
  of Process 1 [rule 4b] as {\it subjective} appearances to such
  observers, thus placing the theory in correspondence with experience.
  We are then led to the novel situation in which the formal theory is
  objectively continuous and causal, while subjectively discontinuous
  and probabilistic.  While this point of view thus shall ultimately
  justify our use of the statistical assertions of the orthodox view, it
  enables us to do so in a logically consistent manner, allowing for the
  existence of other observers."\\
  \ldots\ \\
  Everett's goal then was to show that the memory records of an
  observer as described by his no-collapse theory would match those
  predicted by the standard theory with the collapse dynamics.  The main
  problem in understanding what Everett had in mind is in figuring out
  how this correspondence between the predictions of the two theories
  was supposed to work.\\
  \ldots\ \\
  {\bf Many Worlds}\\
  DeWitt's many-worlds interpretation is easily the most popular reading
  of Everett.  On this theory there is one world corresponding to each
  term in the expansion of {\bf E} when written in
  the preferred basis (there are always many ways one might write the
  quantum-mechanical state of a system as the sum of vectors in the
  Hilbert space; in choosing a preferred basis, one chooses a single set
  of vectors that can be used to represent a state and thus one chooses
  a single {\it preferred} way of representing a state as the sum of
  vectors in the Hilbert space).  The theory's preferred basis is chosen
  so that each term in the expansion of {\bf E}
  describes a world where there is a determinate measurement record.
  Given the preferred basis (surreptitiously) chosen above,
  {\bf E} describes two worlds: one where $J$
  (or perhaps better $J1$) determinately records the measurement
  result "spin up" and another where $J$ (or $J2$)
  determinately records "spin down".

  DeWitt and Graham describe their reading of Everett as follows:
  "[Everett's interpretation of quantum mechanics] denies the existence
  of a separate classical realm and asserts that it makes sense to talk
  about a state vector for the whole universe.  This state vector never
  collapses and hence reality as a whole is rigorously deterministic.
  This reality, which is described {\it jointly} by the dynamical
  variables and the state vector, is not the reality we customarily
  think of, but is a reality composed of many worlds.  By virtue of the
  temporal development of the dynamical variables the state vector
  decomposes naturally into orthogonal vectors, reflecting a continual
  splitting of the universe into a multitude of mutually unobservable
  but equally real worlds, in each of which every good measurement has
  yielded a definite result and in most of which the familiar
  statistical quantum laws hold."\\
  \ldots\ \\
  Perhaps more serious, in order to explain our determinate
  measurement records, the theory requires one to choose a preferred
  basis so that observers have determinate records (or better,
  determinate {\it experiences}) in each term of the
  quantum-mechanical state as expressed in this basis. 
\end{quote}

These quotes from Barrett's 1998 article in the Stanford Encyclopedia 
of Philosophy emphasize the central importance of specifying a basis, and 
it gives many references to fill in and back up the description it gives of 
the contemporary situation. However, that review  does not mention David 
Deutsch's ambitious and detailed 1985 attempt\cite{dd85} to solve the 
basis problem. That omission is justified by the fact that Deutsch's approach 
was found by Sara Foster and Harvey Brown\cite{fb} to be marred by many 
logical flaws. But in spite of these flaws, it will 
be useful to describe  Deutsch's 1985 attempt in order to get an idea of 
the problems involved in the construction of a many-worlds interpretation.

{\bf 2. Deutsch's 1985 Proposed Solution of the Many-Worlds Basis Problem.} 

Deutsch clearly recognized that a key problem in creating
a many-worlds interpretation is the problem of defining the preferred
basis. He calls it the `interpretation basis', and his forty-page paper is 
devoted largely to task of defining it. The other main problem is to 
understand how a relative probabilility comes to be associated 
with each of the alternative possibilities, all of which occur jointly 
and concurrently in the evolving quantum state. Since the occurrence of 
any one of the possible outcomes is accompanied by the occurrence of each 
of the other possible outcomes it is a problem to understand, solely on 
the basis of the Schroedinger equation alone, how one of the possibilities 
can be, say, a billion times less likely to appear than some other one. 
I shall not delve here into this interesting problem, but will focus 
on the basis problem.

Solving the basis problem seems to demand mathematically
detailed rules that specify (1), when the full reality splits into 
these distinct worlds, or branches, (2), what these separate parts are, 
and (3), which variables associated with these different worlds, or branches,
acquire, or possess, definite values in each such world, or branch.
Deutsch recognized the three aspects of the problem, and tried to
provide well defined answers to each.

Of course, every interpretation of quantum theory that goes beyond the 
pragmatic Copenhagen approach of merely providing a set of rules the connect
human experiences must provide analogous kinds of rules. But the mathematical 
requirements on the structure of these rules is far more stringent for 
a many-world theory because such a theory allows no collapse, and everything
must be determined exclusively in terms of the one basic reality, which
is the one quantum state of the universe that evolves in accordance with
the Schroedinger equation: no other process is allowed. But as 
Foster and Brown emphasize, in {\it collapse} theories the specification of 
these rules "need only be within the context of {\it some} theory, not 
necessarily quantum theory." The encompassing theory could allow, for example,
as von Neumann's theory does, experiential realities to enter dynamically 
into the determination of the basis, instead of being mere consequences of 
some choice of basis that is itself wholly determined by the unitary 
evolution of the quantum state in accordance with the Schroedinger equation. 

Deutsch recognized and stressed the critical importance of this basis problem,
and made a determined effort to solved it: 

"The remedy which I propose in section 7 is that an extra rule
be appended to the quantum formalism, specifying how the interpretational 
basis depends at each instant of time on the physical state and dynamical 
evolution of a quantum system."  

The first problem that he tries to resolve is what he calls "the notorious 
difficulty of specifying the preferred instants $t$ \ldots\ when a 
measurement is completed."  These instants must be defined solely in terms 
of (1), the single evolving state of the universe, which evolves always via 
the Schroedinger equation, and (2), the law that governs
this continuous unitary evolution. 

The second problem is that at each of these distinct ``instants of 
splitting'' the unified actual universe must be decomposed in a 
specified way into two systems, the `observed system' and 
`observing system'. Hence the rule that defines how the unified universe 
is to be split at that instant into these two special parts must be 
specified in terms solely of the state vector itself and its law of 
evolution. 

The third problem is to define for each instant a set of evolving basis 
vectors so that the state of the universe can at every instant be decomposed 
in a unique way into a sum of component vectors that can be associated with 
different worlds, or branches of reality, that possess definite values for 
the quantities ``measured'' at each instant of splitting. 

Deutsch solves the problem of dividing the unified physical world
into the well defined  physical parts that will ultimately 
acquire `possessed values' by restricting his considerations to a model 
world that has only a {\it finite} set of states, and dividing this set of 
states in some way into subsystems each having a `prime number' 
of states in each subsystem. One evident problem with that approach 
is the essential role played by this finiteness assumption and the
`prime factor' rule: Deutsch does not indicate how this `prime number rule', 
which is purely kinematic (it does not involve the equations of motion)
will carry over to real systems, in which every particle carries with 
it an infinite number of states, and in particular to the systems 
that carry the stored memories associated with our experiences. These
systems are presumably made of many particles, and are strongly dependent 
on the dynamics. 

Once the separation of the universe into the observed and observing systems
is made, Deutsch specifies the instant at which the splitting occurs
to be essentially the instant at which the interaction between observed and
observing system vanishes. But the ubiquitous tails of real systems would
make that condition difficult to satisfy in the physical universe.

Once the instants of splittings are fixed, and the separation of the universe
at each of these instants into the observed and observing system is specified,
one needs to define, for {\it every} instant of time, the preferred basis. 
This basis defines the decomposition of the quantum state of 
the universe into a unique sum of vectors each of which is parallel to one 
of the vectors in the preferred set of basis vectors. Each of these particular
component vectors is supposed to correspond to a separate world, or branch
of the universe. The decomposition must be such that if a probability is 
assigned to each such world in accordance with usual rule, which equates 
it to the square of the length of the corresponding vector, then the 
expectation values of certain special quantities, namely those represented 
by {\it operators whose eigenvectors are these basis vectors}, are required 
to agree with the general quantum mechanical rule for expectation values. 
This is how Deutsch proposed to endow the evolving non-collapsing quantum 
universe with some definite outcomes in a way concordant with the quantum 
probability rules: each individual branch would have `possessed values' 
appropriate to that branch, and the branch would be given the appropriate
statistical weight. But the entire construction depends crucially on the 
idea that a particular well-defined set of preferred basis states is 
specified by the evolving non-collapsing quantum state of the universe.

Deutsch tried to deduce the structure of this preferred basis set by 
imposing the requirement of no-faster-than-light-influence, but his 
argument is, as was correctly noted by Foster and Brown, logically flawed. 
The primary condition on a many-world interpretation is that there be no 
actual collapse: the {\it values} associated with the preferred basis 
vectors are supposed to be "possessed" by the systems, or to exist within 
nature, without there being any collapse that actually eliminates the 
cross terms. However, the crucial equation (Eq. 50) that gives the structure 
of the preferred basis does not follow from Deutsch's argument, insofar as 
the no collapse condition is maintained. The result {\it would} 
follow if the appropriate collapse were to occur. That is, Deutsch's proof 
would be valid if he were trying to define the preferred basis in the 
context of a collapse-type theory, but it is not valid in the many-worlds 
context. This error is not just a small slip-up that can by fixed by a minor
correction. Foster and Brown emphasize this problem, and also problems
with the existence and uniqueness of solutions to the conditions that
Deutsch proposes. Michael Lockwood reports in his 1989 book
{\it Mind, Brain, and the Quantum,} (p. 236) that Deutsch 
"has failed so far to find any plausible way to meet Foster 
and Brown's objections." Lockwood's own solution brings in "Mind" in an 
important way, and is thus not a "Many-Worlds Interpretation" in the
DeWitt-Deutsch sense of this term.

{\bf 3. The Core Basis Problem}

This section identifies the core basis problem. 

The essential point is that if the universe has been evolving since the 
big bang in accordance with the Schroedinger equation, then it must by now 
be an amorphous structure in which every device is a smeared-out cloud 
of a continuum of different possibilities. Indeed, the planet earth would
not have a well defined location, nor would the rivers and oceans, nor the 
cities built on their banks.  Due to the uncertainty principle, each particle 
would have had a tendency to spread out. Thus various particles with various 
momenta would have been able to combine and condense in myriads of ways into 
bound structures, including measuring devices, whose centers, orientations, 
and fine details would necessarily be smeared out over continua of 
possibilities. The quantum state would be, to first order, a superposition 
of a continuum of slightly differing classical-type worlds with, in 
particular, each measuring device, and also each observing brain, smeared 
out over a continuum of locations, orientations, and detailed structures. 
But the normal rules for extracting well defined probabilities from a 
quantum state require the specification, or singling out, of a 
{\it discrete set} (i.e., a denumerable set) of orthogonal subspaces, one 
for each of a set of alternative possible experientially distinguishable 
observations.  But how can a particular discrete set of orthogonal 
subspaces be picked out from an amorphous continuum by the action of 
the Schroedinger equation alone? 

For example, if one has in the wave function of the universe,
in addition to the contribution from some particular wave function
of a Stern-Gerlach device, also contributions from all slight deviations
of this wave functions from this original form, then this collection, if no 
one deviation is singled out from its immediate neighbors in some way, 
defines no proper subspace: any collection of vectors in a full neighborhood 
of a vector spans the entire vector space. This fact poses a problem in 
principle for any deduction of probabilities from the Schroedinger
dynamics alone: how can a specific set discrete orthonormal subspaces be 
specified by the continuous action of the Schroedinger equation on a 
continuously smeared out amorphous state?

The essential role of the human participant/observer in the
orthodox Copenhagen interpretation is to pick out a particular discrete 
subspace (or a particular set of orthogonal subspaces) from a continuum of 
logically possible alternatives. He is able to do this by not being part
of the universe that evolves via the Schroedinger equation. 
The fact that the participant/observer who chooses the experiment, and  
observes the outcome, stands, in this sense,  outside the quantum universe, 
but acts upon it and observes it, is a key logical aspect of the orthodox 
interpretation. In the von Neumann version this choice is achieved by the 
notorious ``process I'', which likewise is not determined by the unitary 
Schroedinger process, but specifies a particular way of dividing the full 
Hilbert space into a {\it denumerable} set of orthogonal subspaces.

Any theory that proposes that the entire universe be controlled by 
the Schroedinger equation alone must explain how that continuous 
unitary process can pick out a particular denumerable set of `observable' 
orthogonal subspaces with nonzero probabilities, in such a way that these 
nonzero probabilities become associated with empirically
experienced alternatives.  

Most contemporary efforts to solve the ``measurement'' problem in quantum
theory rest heavily upon ``environmental decoherence effects.'' 
One of the effects of this interaction of a system with its environment is 
to convert what would otherwise be a wave packet that can extend coherently 
over a large distance, in various dimensions, into what is essentially a 
{\it mixture} of narrow gaussian wave packets. The mixture is, essentially, 
a sum over a {\it continuum} of displaced narrow Gaussians in the variable 
associated with the wave function \cite{zeh}. These Gaussian states are 
dynamically robust under the effects of interactions with the environment,
and have classical-type properties that make then good candidates for
the states that correspond to definite experiences. 

Specifically,  the effect of environmental decoherence is to
reduce a typical state $\Psi(x')$ to a mixture of states of the form
$$ 
\Psi_x(x') = N(x) \exp -a^2(x'-x)^2,    \eqno (3.1)
$$
where $N(x)$ is a weight factor, and the subscript $x$, which 
labels the center-point, or peak, of the Gaussian wave packet, ranges over 
a {\it continuum} of values. None of these states is orthogonal to any other.
(Here, to make the idea clear, I am exhibiting only one single continuum 
variable, $x$. But all position-type variables would be affected in a 
similar way.)  

This decoherence mechanism eliminates certain interference effects, but it
does not solve the basis problem.  There will be a {\it continuum} of these 
Gaussians, (3.1), and they {\it overlap}: i.e., they are not orthogonal. 
As before, if one takes a set of these state consisting of one of them 
together with those displaced from it by a distance $\Delta x$ less than 
any epsilon greater than zero then this set will included a {\it continuum} 
of states that span a large part of the Hilbert space of functions of the 
variable $x'$ (See \cite{kl}).  Since no one of this continuum of states 
is singled out from the rest by the Schroedinger dynamics, and they jointly 
span a large part of the Hilbert space in question, one does not 
immediately obtain from the Schroedinger dynamics plus decoherence the 
needed denumerable set \{$P_i$\} of orthogonal projection operators, or 
the denumerable set of orthogonal subspaces that they specify.

A natural candidate for the needed denumerable set of orthogonal subspaces 
associated with the continuum of states (3.1) is the set of subspaces defined
by the different eigenvalues of the reduced density matrix associated with 
that mixture of states. That candidate is, in fact, the one chosen by Deutsch 
in his 1985 paper. However, that choice is not satifactory. One wants states 
that are highly localized, like a classical state. But the eigenstates of 
the reduced density matrix associated with the states (3.1), i.e., the 
eigenstates of a density matrix of the form 
$$
<x'|\rho_r |x''> = N(x',x'') \exp -a^2(x'-x'')^2/2,   \eqno (3.2)
$$
with a slowly changing normalization function $N(x',x'')$,  are spread out 
over large distances, rather than being compressed into small regions, like 
the states (3.1) themselves are. Taking the eigenstates of the reduced 
density matrix goes in the wrong direction. Because this natural choice is 
unsatisfactory one is left with the core problem of how to specify, without 
appeal to any process except the continuous Schroedinger unitary evolution, 
the denumerable set of orthogonal subspaces that are to be associated 
with definite observed values and finite probabilities.

{\bf 4. Deutsch's Multiverse Proposal}

Recently Deutsch has put forward a new many-worlds concept
labelled the "multiverse". In the popular account given in his book
{\it The Fabric of Reality} the multiverse would appear to be a collection
of classical-type universes each one similar to the single quasi-classical
world introduced by David Bohm in his famous pilot-wave model. But that 
cannot be Deutsch's real meaning because Bohm's classical-type universe is
governed, in part, by nonlocal forces, and a key virtue of the many-worlds 
approach, often emphasized by Deutsch, is that it does not involve or 
entail any kind of faster-than-light transfer of information. 

In his recent paper {\it The Structure of the Multiverse} \cite{dd01}
Deutsch gives a more detailed glimpse into his thinking, 
and he emphasizes at the outset the existence of the basis problem, and 
the demand for no faster-than-light transfer of information. Deutsch's 
new proposal involves different quasi-classical aspects of the quantum 
universe that persist and evolve in parallel over finite temporal intervals 
but can eventually {\it recombine} to produce quantum interference effects. 
This idea is much closer to standard ideas than Bohm's model, in which the 
quasi-classical world persists for all time, and quantum interference
effects are explained by a special nonlocal force, rather than by the merging
of previously distinct quasi-classical components.
  
In his new work Deutsch again emphasizes the basis problem, and I think 
it is  fair to say that a many-worlds interpretation is not defined
until it solves in principle the basis problem. Indeed, the basis problem 
is {\it the} problem that any interpretation of quantum theory must resolve 
in some way. Thus the central idea of the Copenhagen interpretation was 
to imbed the quantum system in a larger system that specifies the 
preferred basis by bringing in "measuring devices" that are set in place 
by a classically conceived process. In von Neumann's formulation there is 
the infamous ``Process I", which likewise lies outside the process governed
by the Schroedinger equation. The key challenge that must be met by
a satisfactory many-worlds interpretation is to solve this basis
problem within quantum theory itself, without bringing in any
outside dynamical process.  

Deutsch's program is to create a satisfactory many-worlds-type theory
by considering nature to be a quantum computer. Deutsch cites Feynman's 
paper\cite{rf} on quantum mechanical computers. In this paper
Feynman says:
\begin{quote}
Of course, there are external things involved in making the measurements
and determining all this [setting up the computer], which are not parts
of our computer. Surely a computer has eventually to be in interaction
with the external world, both for putting in data and for taking it out.
\end{quote}
Here Feynman is following the orthodox idea that a computer is a quantum
system, and as such must be considered imbedded in a larger world in which
we prepare initial conditions, including the structures of all devices,
and then extract outcomes. Deutsch's problem is, accordingly, to show how by 
considering the whole universe to be a quantum computer one can solve
this heretofore unresolved interpretational problem that has beset those 
trying to create a satisfactory many-worlds theory.

To emphasize the problems that others have encountered I quote some
passsages from Zurek's 1998 review\cite{z} of the status 
of attempts to use environmental decoherence (abetted, if helpful, by 
coherent histories) to solve the interpretational problem.
\begin{quote}  
It will be emphasized that while significant progress has been made \ldots\
much remains to be done on several fronts which all have implications on the
overarching question of interpretation. 
\ldots\ \\
Thus, while what follows is perhaps the most complete discussion of the 
interpretation implied by decoherence, it is still only a report of 
partial progress.\\
\ldots\ \\
The advantages of Everett's original vision was to reinstate quantum theory
as a key tool in search of its own interpretation. The disadvantages
(which were realized only some years later, after the original proposal 
became more widely known) include (i) the ambiguity in what constitutes
the branches \ldots\ (ii) re-emergence of the questions about the origin
of probabilities \ldots\ (iii) it was  never clear how to reconcile 
unique experiences of observers with the multitude of alternatives present
in the MWI wave function.\\
\ldots\ \\
The interpretation based on ideas of decoherence and [environmentally
induced] einselection has not really been spelled out to date in any detail.
I have made a few half-hearted attempts in this direction, but, frankly, 
I was hoping to postpone this task, since the  ultimate questions tend to
involve such `anthropic' attributes of the `observership' as `perception',
`awareness' or `consciousness', \\
\ldots\ \\
Einselection chooses a preferred basis \ldots\. It will often turn out
that it is overcomplete. Its states may not be orthogonal, and, hence,
they could never follow from the diagonalization of the density matrix.

\end{quote}
These quotations show that Zurek recognizes the core continuum problem
that I have been describing.

How then does Deutsch deal with this basic problem, which is that a 
quantum universe evolving via the Schroedinger equation does not seem to be 
able to pick out a {\it discrete set} of preferred directions (or subspaces) 
in Hilbert space that specify quasi-classical properties that can be assigned 
definite ``possessed values'' with well defined probabilities.

His approach is first to define, in terms of his quantum-computer
formulation of quantum theory, certain mathematical requirements for 
a portion of the full quantum computer to be `behaving classically'.
These requirements depend upon certain directions in Hilbert space
being singled out as the ones that correspond to `possessed values'.
Those special directions are not specified beforehand, but are defined by 
the dynamics, in the sense that {\it if} with some choice of these directions
a certain system behaves classically then these special directions are picked
out by that classicality condition itself. Thus the preferred basis is defined
dynamically, by a classicality condition.

This is eminently reasonable, as far as it goes. By it does not address
the basic continuum problem, which would be to show that the classicality
conditions specify, at each moment at which ``possessed values'' are defined, 
a {\it denumerable set} of mutually orthogonal subspaces within the full 
amorphous many-worlds universe, in which there will be continua of
neighboring environmentally robust configurations in which the detectors 
that define the preferred axes of his qubit-measuring devices are slightly 
re-oriented. The basic problem is to get out of the {\it continua} specified by
the Schroedinger dynamics plus environmental decoherence the {\it denumerable}
sets of orthogonal subspaces needed to make the probability formulas work. 
This core issue was not addressed.

{\bf 5. Zurek's Approach and Records}

Deutsch's multiverse theory is in general accord with Zurek's
approach\cite{z}, and in the end Deutsch appeals to Zurek's
arguments to support the idea that the `classically behaving'
aspects of the quantum universe will explain the `classical world' 
in which we live and act.

Zurek's treatment is wrapped up, as it should be, with his treatment of 
probabilities. In this connection he says:
\begin{quote}
A true `frequency' with a classical interpretation cannot be defined at 
a level which does not allow `events'---quantum evolution which leads to 
objectively existing states---to be associated with the individual
members of that ensemble.
\ldots\
The reduced density matrix $\rho$, which emerges following the interaction
with the environment, and a partial trace will always be diagonal in the 
{\it same} basis of einselected pointer states $\{|i>\}$. These states help 
define elementary `events'. Probabilities of such events can be inferred 
from their coefficients in ${\rho}$, which have the desired `Born rule'
form. 
\end{quote}
Zurek's discussions of `existing' values and probabilities
rests on this model, in which the reduced density
matrix is diagonal in the {\it same} states that specify the values
that are supposed to `exist'. But that model must come to grips with
the continuum problem, which is emphasized elsewhere in his paper by 
statements such as:
\begin{quote}
In that case [which in fact arises in the quantum field theoretic treatment
of the measurement process] the states which are most immune to 
decoherence in the long run turn out to be \ldots\ Gaussians. \ldots\
Hence coherent states [which are Gaussians] are favored by decoherence.
\ldots\
Einselection chooses a preferred basis in the Hilbert space in recognition
of its predictability. That basis will be determined by the dynamics of the
open system in the presence of environmental monitoring. It will often turn 
out that it is over complete. [This is precisely the case for the Gaussian's 
mentioned above.] Its states may not be orthogonal, and, hence, they would 
never follow from diagonalization of the density matrix. 
\end{quote}

Thus Zurek stresses that the robust Gaussian-type 
states picked out by environmental decoherence belong to 
a {\it continuum of non-orthogonal states}, and hence do not generally 
define a denumerable set of orthogonal subspaces, and are not the 
basis states defined by diagonalization of the reduced density 
matrix. This leads to the question: Does the diagonalization of the 
reduced density matrix yield physically appropriate basis states?

A partial answer this question is obtained by noting that 
for constant $\rho$ the reduced density matrix is essentially
$$
<x'|\rho_r|x''>= 
\sqrt{2/\pi}\int dx \exp -(x'-x)^2 <x'|\rho|x''> \exp -(x''-x)^2 \eqno (5.1)
$$
$$
= \exp -(1/2)(x'-x'')^2 <x'|\rho |x''>    \eqno (5.2)
$$

The eigenfunctions of this $\rho_r$ are the functions 
$\exp -ipx$, which are spread uniformly over all space. (To get normalized 
functions one can formulate the problem with periodicity over a large 
interval.) Thus in this simplest case the diagonalization effectively 
undoes the localization effect produced by the decoherence: it does not
produce localized, robust, quasi-classical states.

Zurek emphasizes, as did Everett, the critical importance 
of {\it records.} A record is typically stored in the state of a 
{\it relative coordinate} in which the potential energy has two deep, 
well separated wells. It has been suggested, in this case, that if the
energy gap between, on the one hand, the two almost degenerate states 
of lowest energy and, on the other hand, the higher-energy 
double-well energy eigenstates is large compared to the energy of 
the pertinent environmental states then those higher-energy states can 
be ignored. The problem then reduces to one in which only the two 
lowest energy states are considered, together with the states of the 
environment. The reduced density matrix would be a $2 \times2 $ 
matrix that, apart from an overall normalization factor, has $(1+b, 1-b)$ 
on the diagonal and the value $a$ in the two off diagonal slots, with $a$ 
generally much smaller than $b$. The two eigenstates of this reduced 
density matrix will, up to corrections of order $a/b$, each be concentrated 
in one well or the other, which is the desired result.  

This approximation is essentially equivalent to concentrating the coordinate
space wave function at two points, and neglecting the connecting network
of coordinate-space points. However, there is a qualitative difference
between setting the other elements to zero and taking them to be very tiny.

To illustrate this point consider first the easily solvable case of the 
$3\times 3$ matrix with diagonal elements $(1,c,1)$, next-to-diagonal 
elements $a>0$, and zeros in the two corners.  The eigenvalues are 
$(1,(1+c+r)/2,(1+c-r)/2)$, with $r=\surd ((1-c)^2 +8a^2)$, and the 
eigenvectors are $(1, 0,-1)$, $(1, s+, 1)$ and $(1, s-, 1)$, with 
$s\pm = 4a/(1-c \pm r)$. For $a$ and $c$ very small, but nonzero, one 
has $s+ \approx 2a$, and $s- \approx -1/a$, and hence the triad of 
eigenstates is unlike the triad of localized states 
$$
(1,0,0),(0,1,0),(0,0,1),   \eqno (5.4)
$$  
which are possible eigenstates for $a=0$.

This example illustrates the general fact that for almost any matrix 
that is symmetric along the diagonal [i.e., with $<i|M|j> = <-i|M|-j>$, 
with the states labeled $(n,n-1, ...,-n)$], and that has $<i|M|j> \neq 0$  
for all pairs $(i.j)$, all of the eigenstatates will be either symmetrical 
or anti-symmetrical: this property of our $3\times 3$ example carries over 
to the general symmetric (along the diagonal direction) case. [One can prove 
this result by making the anzatz that the eigenstates are either symmetric 
or antisymmetric [i.e., $<i|\phi> = \pm <-i|\phi>$] and noting 
that the equations for the symmetric and antisymmetric eigenvectors have 
$n+1$ and $n$ solutions, respectively,  and that equations for the 
symmetric and antisymmetric eigenfunctions are different.] 

This result for the exactly symmetrical case carries over to the
nearby almost-symmetrical cases by analyticity. But the general rule 
for specifying the preferred basis must cover the special case of the 
{\it nearly}  symmetrical potential wells. Since the diagonalization of 
the reduced density matrix does not lead to the correct (i.e., localized) 
eigenstates in the nearly symmetrical cases it cannot be the correct 
general rule. Thus although the diagonalization of the reduced density 
matrix does define a denumerable set of orthogonal subspaces, it cannot 
always do the job that the participant/observer does in orthodox 
Copenhagen quantum theory, and that ``process I'' does in von Neumann 
quantum theory, which is to pick out a discrete set of orthogonal 
subspaces that are associable with alternative possible experiences.

The example of the nearly symmetrical case shows that the diagonalization
of a reduced density matrix does not define the needed discrete orthogonal 
states. Robustness is the essential quality. More specifically, it is the 
quality of a ``record'' not only to endure, but to be creatable, 
reproducible, and able to control essentially classical behaviors without 
being appreciably degraded.\cite{hps} The effects of the environment are 
important in this connection, because they destroy the possibility that 
certain states could be records. Thus environmental decoherence does have 
powerful effects. But that does not solve the problem of extracting, by means 
of the Schroedinger equation alone --- from the amorphous state that is 
created by the evolution of the big bang state via the Schroedinger equation 
--- a discrete set of orthogonal basis record-type states. In this 
Schroedinger state every atom, and conglomeration of atoms, will be a 
smeared out continuum. The essential reason why the founders had to regard 
the quantum system as being imbedded in an enveloping process, and why von 
Neumann had to introduce his Process I, was to solve this problem of specifying
a discrete basis for the probability computations associated with our
apparently discrete (Yes or No) experiences, when the Schroedinger equation 
generates continuous evolution of an ever-amorphous state. Specifing a 
discrete set of basis states solely from the continuous structure 
generated by the Schroedinger equation would appear to be an impossible 
feat. In any case, no claim to having constructed a rationally coherent 
many worlds/minds formulation of quantum theory is justified unless this 
discrete basis problem is clearly recognized and adequately resolved.

\newpage
\noindent {\bf Acknowedgements}\\
Communications from Professors H. D. Zeh, E. Joos, and 
W. H. Zurek have contributed significantly to the form and content
of this paper.\\

\end{document}